%% file: main.tex
\documentclass[final,5p,times,twocolumn]{elsarticle}

\usepackage{graphicx}
\usepackage{amssymb, amsmath, amsfonts}
\usepackage{wasysym}
\usepackage{color}

\newcommand{\versor}[1]{\boldsymbol{\hat{#1}}}
\newcommand{\vect}[1]{\boldsymbol{#1}}
\newcommand{\mat}{\begin{bmatrix}}
\newcommand{\matf}{\end{bmatrix}}

\journal{Nature Geoscience}

\begin{document}

\begin{frontmatter}


\title{The Sands of Phobos: The Martian moon's eccentric orbit refreshes its surface}



\author[1,2]{Ronald-Louis Ballouz*}
\ead{baru-zu.ronarudo@mmx.jaxa.jp}
\author[1]{Nicola Baresi}
\author[1]{Sarah T. Crites} 
\author[1]{Yasuhiro Kawakatsu}
\author[1]{Masaki Fujimoto}
         
\address[1]{Institute of Space and Astronautical Science, Japanese Aerospace Exploration Agency, Sagamihara, Japan}
\address[2]{Department of Astronomy, University of Maryland, College Park, MD, USA}


\end{frontmatter}

\input{abstract}

\section{Introduction}
\label{S:1}

Previous dynamical analyses of Phobos' decaying orbit have shown that changes to its tidal environment lead to surface mobility and evolution, such as the formation of mass-wasting features \cite{Shi2016} and linear grooves \cite{Hurford2016}.  These studies ignore the eccentricity of the moon's orbit, as this would have a small effect over the dynamical timescales set by the orbital decay rate of the planetary satellite ($10^8$ years).  Here, we model the effect of Phobos' eccentric orbit at its current distance, and demonstrate that time-varying forces can lead to substantial mass motion and surface erosion over much shorter timescales ($10^3$-$10^6$ years).  We find that librations induced by the eccentricity of the Martian moon can vary dynamical slopes by up to 2$^{\circ}$ per orbital period (7hr 39 min). These variations are sufficient to trigger a slow erosion process in high-slope regions that experience the largest changes in the tidal environment. 

Using direct numerical simulations of particle dynamics, we demonstrate this new mechanism for surface mobility on Phobos for the first time. In addition, we find that these regions of high surface mobility coincide with one of two distinct "color" units revealed by multispectral and hyperspectral imaging \cite{Fraeman2014}: the blue unit.  The blue unit is characterized by a relatively high albedo and increasing visible to infrared continuum slope. In contrast, the red unit is defined by a slightly lower albedo and relatively steeper spectral slope. It has been unclear whether these two geologic units are compositionally different, or if the spectral differences are due to space weathering, as there is an observed lack of strong absorption features in spectral observations of both units\cite{Basilevsky2014, Pieters2014, Murchie2015}. Through a new eccentricity-driven mechanism for surface mobility, we develop a new model for regolith development on Phobos that can explain the relationship of the two color units. We show that this process can transform "red" space-weathered \cite{Hapke2001,Vernazza2009, Kaluna2016} regolith by exposing "blue" sub-surface material in a process analogous to the tidally-induced refreshing of asteroid surfaces \cite{Binzel2010}.

Our finding provides new perspectives on the space weathering process for airless bodies in the Solar System. For example, since the proposed mechanism makes predictions on the rate at which fresh material can be exposed, it is possible to place new constraints on the space weathering timescales of Phobos by comparing current and future observations of its surface. Additionally, a relatively fast surface-refreshing process on Phobos has implications on the origin of the Martian moons.  An accretion scenario (following a giant-impact or during the formation of Mars) or a capture of an inner solar system body implies the presence of mafic materials on Phobos \cite{Fraeman2014, Thomas2011}. Since our analysis suggests that the blue unit represents pristine endogenic material, the observed lack of mafic mineral absorptions in this unit \cite{Fraeman2012} raises challenges for a giant impact scenario. Most of these challenges are likely to remain unsolved until dedicated missions to Phobos and Deimos will visit these remote objects. The Mars Moons eXploration mission\cite{Kawakatsu2017} (MMX), which will visit Phobos and return samples from its surface, will conclusively determine the origin of this enigmatic body and shed light on the history of the Martian system.

\section{Results}
\label{S:2}
We perform a two-step analysis to investigate the dynamical stability of grain particles on the surface of Phobos. In the first step, we model the gravitational attraction of the planetary satellite via a constant density polyhedron model \cite{WernerScheeres1996} and calculate the evolution of its dynamical slopes (Fig.\ 1a) in the framework of the elliptical restricted three body problem (ERTBP, \cite{Broucke1969}). Despite the tiny eccentricity of Phobos' orbit ($e \simeq 0.0151$), this model produces time-varying effects that change the local acceleration across the whole surface of Phobos. In particular, these fluctuations result in dynamical slope changes up to 2$^{\circ}$ per orbit that may trigger mass-wasting events near Stickney and the anti-Mars point (Fig.\ 1b). In the second step, we attempt to quantify the erosion rate in these regions. 
\begin{figure}[h]
\label{fig:GravityVar}
\centering\includegraphics[width=1\linewidth]{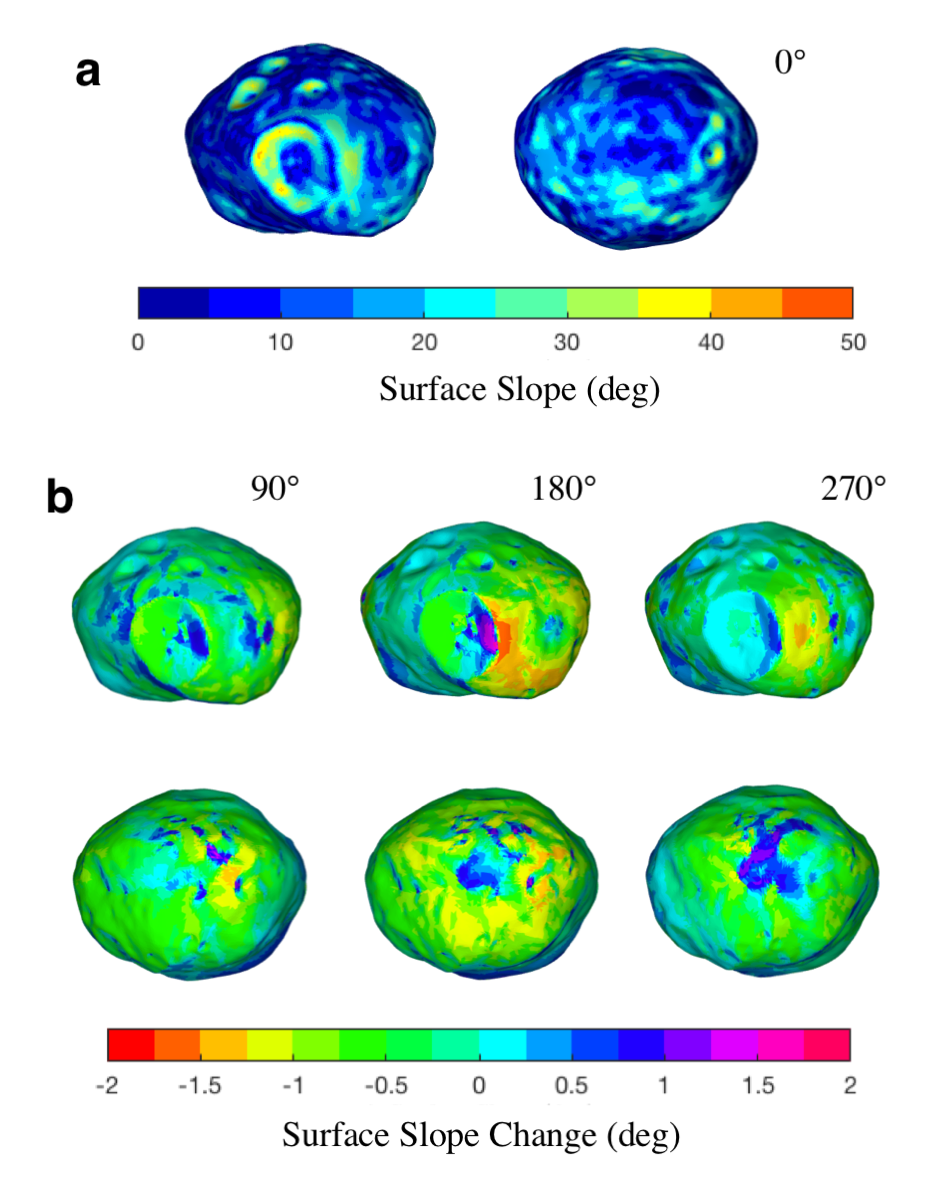}
\caption{\textbf{Changes in the local surface slopes over 1 Phobos orbit.} \textbf{a,} Using a polyhedron shape model of Phobos, we can calculate the surface slopes at its current orbit. \textbf{b,} We evaluate the variations in the surface slope over 1 orbital period. Starting from periaerion, i.e., when the true anomaly, $\nu$, is equal to $0^\circ$, we find that the dynamical slopes of Phobos can vary by up to $\sim$ 2$^{\circ}$ over one orbital period. The largest variations are found in regions along the equator and, especially, in the Mars and anti-Mars directions when $\nu \simeq 180^\circ$.}
\end{figure}

In order to evaluate the magnitude of this effect on transforming Phobos' surface features, we consider the analytical theory of erosion for transport-limited downslope flow\cite{Culling1960}.  For a planetary surface with loose regolith (with little to no cohesion) where the flow of grains is controlled by the transportation rate rather than the regolith supply or production rate, the flow rate, $q$, can be defined as a function of the local surface slope, $\theta)$, as   

\begin{equation}\label{Eq:qtheory}
q = K \tan(\theta),
\end{equation}

where $K$ is the downslope flow constant in units of volume flux per unit time. In previous studies of regolith mobility on small bodies\cite{Richardson2014}, the value of $K$ was estimated for impact-induced disturbances using a Newmark slide-block model\cite{Richardson2005}, and the value of the flow rate was shown to be linear with the slope only when $\theta < 10^{\circ}$ circa \cite{Richardson2014}. For higher slopes, fast granular flows are better modeled as a non-linear function that takes the critical slope,  $\theta_{\mathrm{C}}$, into consideration \cite{Roering1999}.  
Here, we perform local simulations of grain dynamics in order to estimate the value of $K$ for a variety of surface slopes and periodic fluctuations, i.e., $\delta \theta$.  Using a soft-sphere discrete element code, \texttt{PKDGRAV}\cite{ Richardson2011,Schwartz2012}, we set up simulations of a 1.5 m by 1.2 m regolith bed with a porosity of 45\% settled in Phobos gravity ($\simeq 5$ mm/s$^2$) and made up of grains with radii between 1.4 and 1.6 cm. The material properties of the grains are similar to sand of medium hardness\cite{Jiang2015}. The simulated grains have a measured critical angle of repose of $34^{\circ}$. In order to allow the grains to flow freely in the event of an avalanche trigger or creep-induced motion, we impose periodic boundaries in the directions perpendicular to the initial local gravity vector. The simulation setup is illustrated in Fig.\ 2a. 

With the goal of determining a conservative estimate for expected volume flux across different regions of Phobos, we perform simulations for initial surface slopes between 5$^{\circ}$ and 32$^{\circ}$ with $\delta \theta \in [0.5^\circ, 1.5^\circ]$. The highest surface slope achieved across all of the simulations is 34.5$^{\circ}$. The output of the simulations display creeping motion of small grain displacements over the course of a Phobos orbital period. In particular, we find that a granular bed that initially starts at a high slope with significant slope changes will experience the most mass flow. However, even granular beds that are in a sub-critical state and are subject to relatively small variations can lead to surface mobility. We term this a \emph{cold flow} process as opposed to the ``fast" flow of an actual landslide.

The cold flow process is illustrated in Fig.\ 2b with an example of local simulations for the case of a regolith bed that has a mean slope of 31 deg and $\delta \theta$ between $0.625^\circ$ and $1.5^{\circ}$. Here, we highlight the largest particle displacements (anything greater than 3 cm) by showing the initial (black) and final (yellow) positions of the particles along with their tracks (dotted red lines). For each simulation, we measure the total displacement of particles in order to calculate the volume flux across the simulation area over one orbit. For the phase space studied here, which consists of regolith resting at sub-critical angles, we calculate volume flux rates between 10$^{-5}$ to 4 particles/m$^2$/orbit. We show the results of all simulations in Fig.\ 2c. 

\begin{figure}[h]
\label{fig:LocalSims}
\centering\includegraphics[width=1\linewidth]{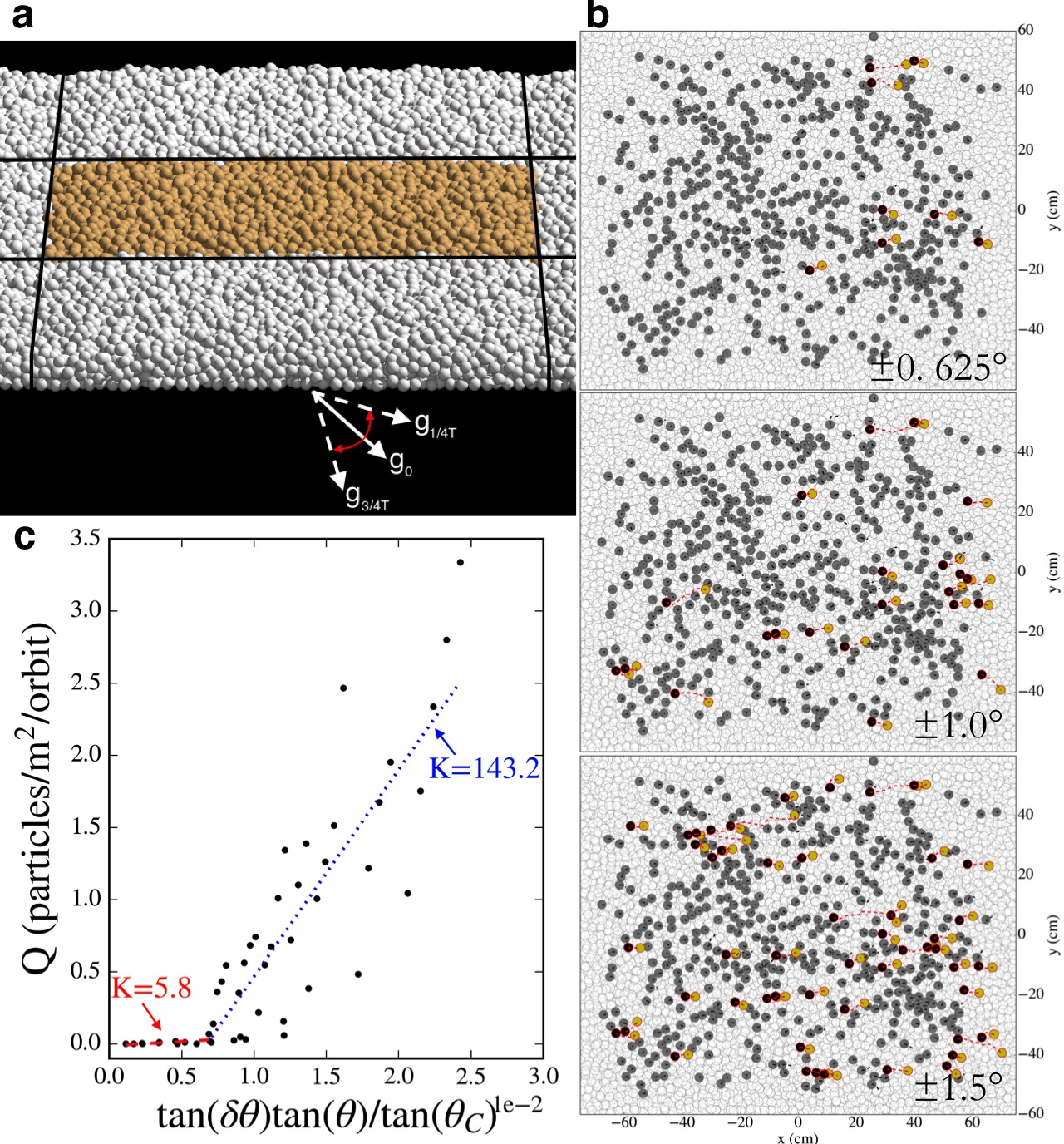}
\caption{\textbf{Local simulations of grain motion on the surface.} \textbf{a,} We set up our simulation with periodic boundary conditions in order to better measure the volume flux induced by Phobos’ eccentric orbit. Arrows show exaggerated variations in the surface slope that are determined by the dynamical calculations demonstrated in Fig.\ 1. \textbf{b,} Comparison of grain motion across three different simulations that have a mean slope of 31$^{\circ}$ and have variable slope change amplitudes as shown in the bottom right corner of each panel. In dark gray, we show the top-most layer of grains. We highlight the largest particle displacements ($> 3$ particle radii) by showing the initial position (black circle) and final position (yellow circle) of these particles, connected by their paths (dashed red lines). \textbf{c,} Aggregating our results across 47 simulations, we formulate a prescription for the expected volume flux $Q$ for a given gravity slope and its variation over 1 orbit, see equations (\ref{Eq:qtheory}) \& (\ref{Eq:Qdata}).}
\end{figure}

We find that across low and high slopes, the expected volume flux due to eccentricity-driven surface slope changes, $Q$, is best described by,

\begin{equation}\label{Eq:Qdata}
Q = K \tan(\theta) \frac{\tan(\delta \theta)}{\tan(\theta_{\mathrm{C}})}.
\end{equation}

Based on the simulations conducted in this study, we find that the magnitude of the particle flow can be described by two linear regimes with different values of the flow constant $K$. The cold flow prescription depends on the proximity of the initial slope to the critical angle of repose and the magnitude of the surface slope-change. We find that the value of $K$ switches from $5.7$ particles/m$^2$/orbit to $143.2$ particles/m$^2$/orbit when $Q/K = 7\times10^{-3}$, which corresponds to a value of $\delta \theta = 0.4^{\circ}$ when $\theta = \theta_{\mathrm{C}}$. Therefore, equation (\ref{Eq:Qdata}) provides a description for the cold-flow mechanism on the surface of Phobos induced by its eccentric orbit. We use this prescription to compare the expected erosion rates in different regions of Phobos with observational data and identify areas where the cold flow mechanism may be dominating the geomorphology of the Martian moon.

Observations by the HiRISE camera on NASA's Mars Reconnaissance Orbiter (MRO) spacecraft and the High Resolution Stereo Camera (HRSC) on ESA's Mars Express spacecraft have produced detailed images showing the heterogeneity of the area in and around Stickney crater (Fig.\ 3a,d). While high albedo features are typically indicative of mass-wasting motion, the area east of Stickney has a relatively smooth surface\cite{Basilevsky2014}, free of streaks that typify granular flow on Phobos \cite{Shi2016}. There have been some suggestions that the color variation in this area is a result of a catastrophic landslide that began in Stickney’s western wall and transitioned to the sub-Martian point by motion of a thin surface layer of material in a long run-out landslide \cite{ShingarevaKuzmin2001, CollinsMelosh2003}. However, these studies either assume small and unrealistic values for the friction properties of the Phobos surface (friction coefficient of 0.01-0.09\cite{ShingarevaKuzmin2001}), or require the development of acoustic fluidization \cite{CollinsMelosh2003}. Another interpretation of this region is that it is asymmetric ejecta deposit from Stickney\cite{Thomas1998}. Here, we offer a novel interpretation.
\begin{figure}[h]
\label{fig:Maps}
\centering\includegraphics[width=1\linewidth]{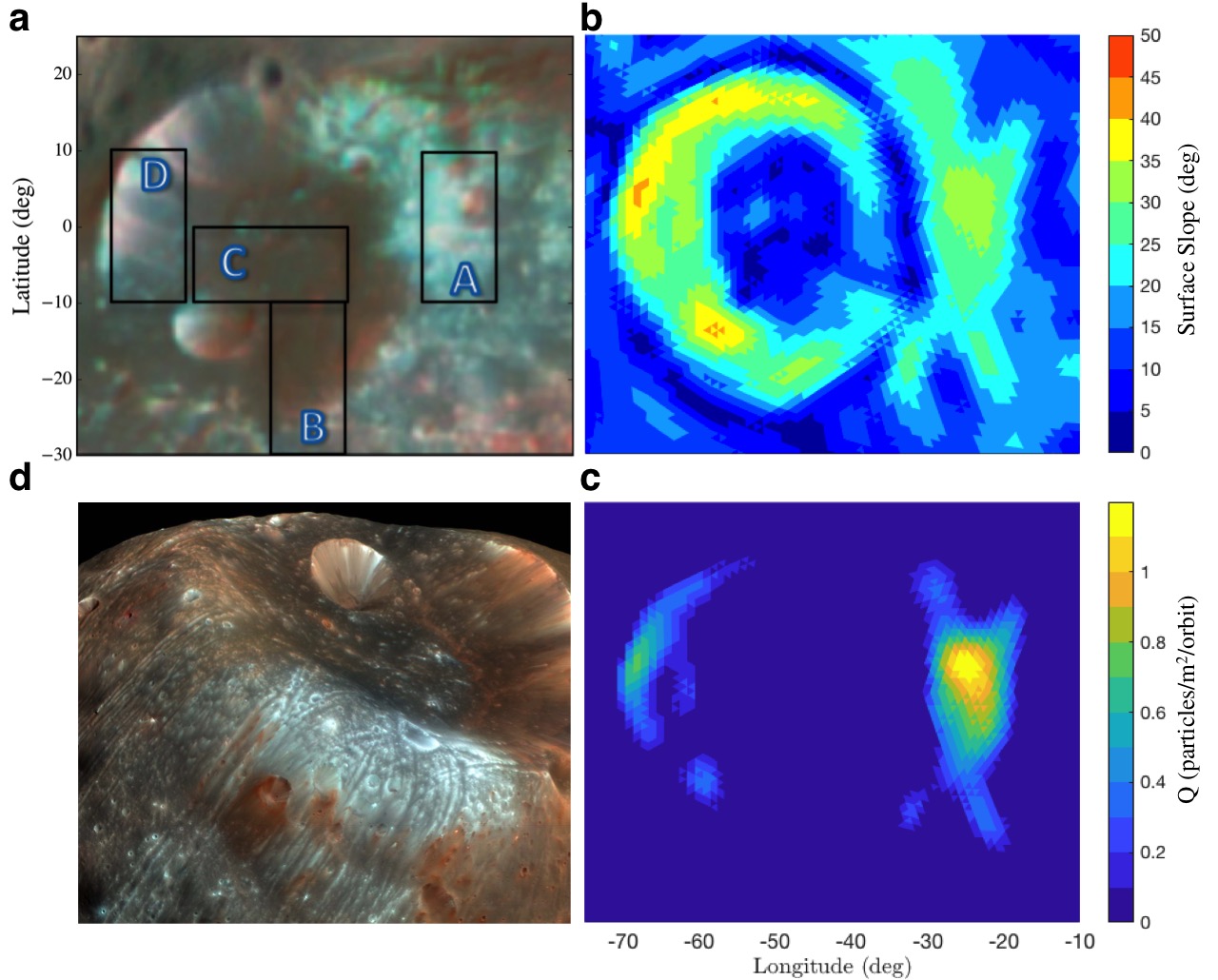}
\caption{\textbf{High-slope regions on Phobos that undergo large variations coincide with blue surface units.} \textbf{a,} Color composite image of Stickney crater derived from Mars Express HRSC data \cite{Patsyn2012}. \textbf{b,} Dynamical slopes in and around Stickney when $\nu = 0^\circ$. \textbf{c,} Highlighting regions of possible mass-wasting based on our estimate of $Q$. \textbf{d,}  Image \texttt{HiRISE\_PSP\_007769\_9010\_IRB} synthesized from three infrared and visible channels showing Stickney crater, highlighting the blue regions east of the crater.}
\end{figure}
By analyzing the spatial distribution of the surface slopes in the region east of Stickney (Fig.\ 3b) and the predicted variation due to Phobos' eccentric orbit, we predict a relatively large amount of regolith flow compared to nearby regions (Fig.\ 3c). This area, labeled A in Fig.\ 3a, should have the most mobile regolith due to the unique combination of steep terrain and high surface-slope changes. In contrast, the relatively steep southern wall of Stickney (labeled B) does not experience large variations in the tidal environment of Mars and--in agreement with our model--shows little indication of blue material (Fig.\ 3d). 

Through this qualitative comparison of Phobos color data with our predicted volume flux, we propose that blue geologic units are composed of sub-surface material that is being continually exposed by the eccentricity-driven librations of Phobos. This suggests that the blue units are the footprints of ``fresh", pristine material, as opposed to the ``old", space weathered regolith that is found in the red unit of the Martian moon. To evaluate this possibility, we compare our model's predicted excavation rate to the likely space weathering rate at Phobos. The results of this analysis are discussed in the next section.


\section{Discussion}
\label{S:3}

In order to determine the efficiency of this process in uncovering fresh material, we consider the possible timescale and depth of space weathering on Phobos. Since we have a poor understanding of Phobos' composition \cite{Murchie2015} and the specific mechanisms that may influence the maturation of its regolith \cite{PietersNoble2016}, we consider a range of space weathering rates to provide some perspective on the possible surface refreshing rate. We use estimates for space-weathering based on previous spectral analysis of main-belt asteroids. Studies on space weathering of S-complex asteroid dynamical families \cite{Vernazza2009, Binzel2010} reveal that space-weathering rate for these bodies is relatively rapid ($< 10^6$ years), suggesting that solar wind ion implantation dominates the process. In contrast, analysis of C-complex asteroids \cite{Kaluna2016} suggests that space weathering is a long term process, with spectral slopes being modified on timescales closer to $10^9$ years. Using the proposed space-weathering rates \cite{Kaluna2016} and slope changes in CRISM reflectance measurements between blue and red geologic units \cite{Fraeman2012}, we estimate that space weathering on Phobos can occur on timescales with a lower limit of 10$^6$ years. 

Due to the lack of in-situ exploration of small-body subsurface properties, we rely on analysis of lunar core samples from the Apollo missions to obtain an estimate on the depth of space-weathered material on Phobos. Measurements of the regolith maturity parameter ($I_s/FeO$ ratio) of 12 Apollo core samples show that the regolith typically transitions between mature weathered material to relatively fresh material around 20-50 cm below the surface \cite{Lucey2006}. Combining this estimate of the depth of weathered material with our assumptions on the space-weathering rate and the predicted excavation rates derived from our numerical simulations, we analyze the efficiency of the eccentricity-driven cold-flow process in exposing fresh material.  We find that, while the calculated excavation rates are relatively small (only a few grains typically move from a 1 m$^2$ high-slope region every Phobos orbit), the accumulated effect over the course of Phobos' geologic time can be quite substantial. In Fig.\ 4, we show the excavation rates for the cold-flow of mm-size particles required to expose material at certain depths (represented by the solid curves) for different timescales. It is important to note that we make a steady-state assumption for regolith as it becomes exposed. In reality, the regolith can mature as it is uncovered. Nevertheless, this analysis provides an order of magnitude estimate of the necessary excavation rates required to expose fresh material.

From our numerical simulations of the cold-flow mechanism, we show that the simulation-derived mean excavation rates (dashed lines) for the labeled regions in Fig.\ 3a. The values of $\theta$ and $\delta \theta$ for these regions are determined based on our dynamical calculation (Fig.\ 1), and are fed into the prescription of mass flux, equation (\ref{Eq:Qdata}). We find that the regions with the highest spatial distribution of blue material in the HRSC data (region A) has the highest mean excavation rate, followed by the western wall of Stickney (region D), and both the southern wall (region B) and the crater floor (region C) have very low predicted excavation rates.
Overall, we find that blue regions represent areas where grain-mobility is high enough to expose fresh material down to depths of $>$ 20 cm at a rate faster than the expected space weathering time-scale (grayed-out region). Red regions have predicted mean excavation rates that may only expose material a few cm deep before $10^6$ years, which is the age where our steady-state assumption may fail, as some asteroids\cite{Binzel2010} are found to rapidly redden in this time-scale. Therefore, we conclude that eccentricity-driven mass-wasting is an efficient process for uncovering pristine un-weathered material on Phobos, represented by blue geologic units found near Phobos’ sub-Mars point. We propose a model for regolith development where i) Phobos is inherently blue, ii) space weathering reddens the surface material, iii) impacts and Phobos' decaying orbit can expose fresh or weathered material depending on crater depth and the characteristic space weathering depth $d_{\mathrm{SW}}$, iv) this freshly exposed material is reddened over time, and v) blue regions are continuously refreshed by mass-wasting that occurs in high-slope regions that experience significant variations in tidal forces due to Phobos’ eccentric orbit.


\begin{figure}[h]
\label{fig:EXR}
\centering\includegraphics[width=1\linewidth]{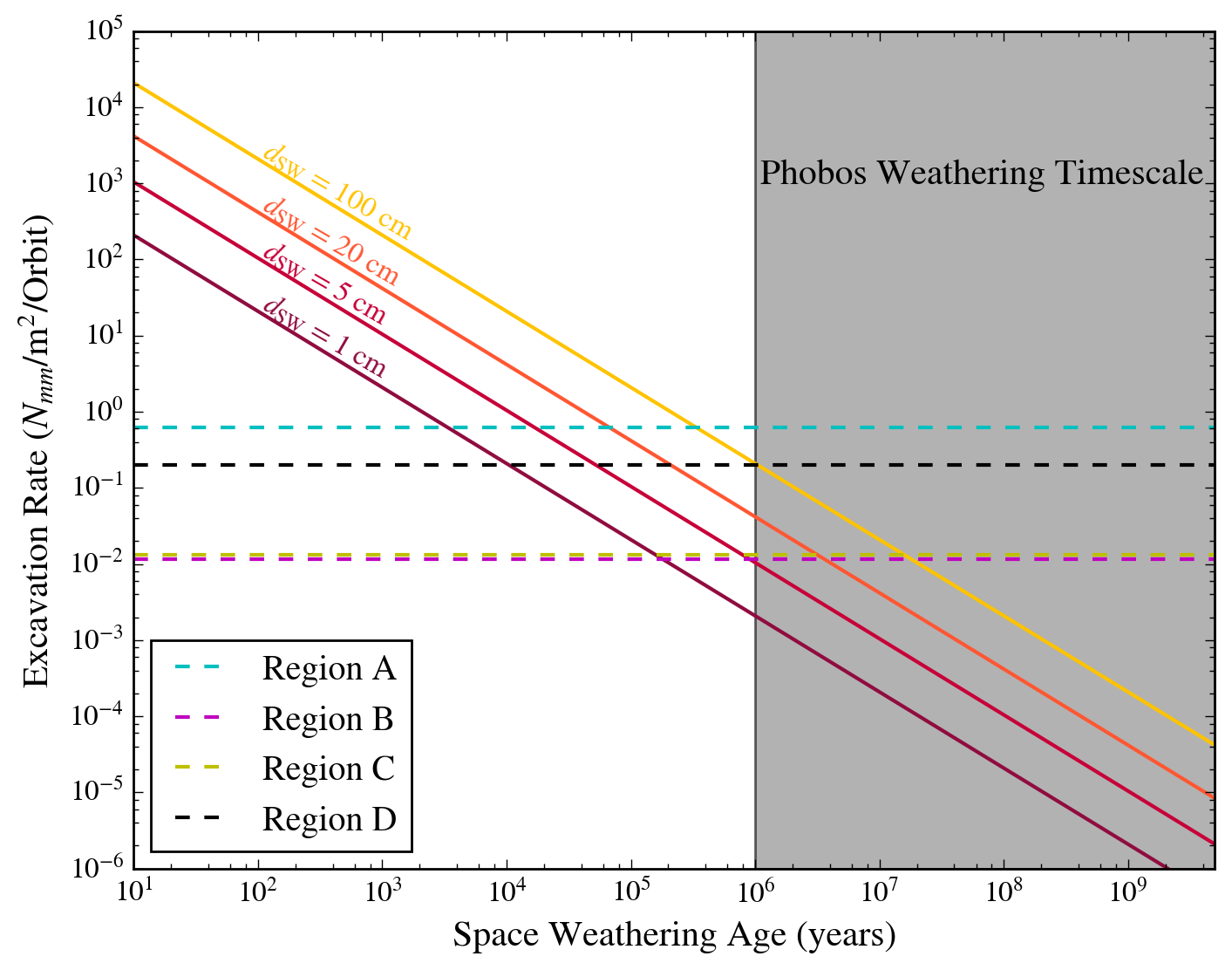}
\caption{\textbf{Required volume fluxes necessary to uncover subsurface material before a certain weathering timescale.} We show the excavation rate required to uncover fresh material buried at different depths ($d_{\mathrm{SW}}$) ranging from 1 to 100 cm that requires a corresponding number of years (Space Weathering Age) to mature. This may be interpreted as the number of mm-size particles, $N_{mm}$, in the top layer of regolith that need to flow out of a 1 m$^2$ region every Phobos orbit (vertical axis) in order to excavate un-weathered material, which takes a given number of years to weather, $t_{\mathrm{SW}}$, down to a depth, $d_{\mathrm{SW}}$, described by the solid curves. The gray area represents the estimated timescale for material on Phobos to mature based on analysis of spectral slope reddening in CRISM data. The dashed lines represent the mean excavation rates of the regions highlighted in Fig.\ 3a: region A (cyan),  region B (magenta), region C (yellow), region D (black). This demonstrates the agreement between blue regions in the HRSC color data and the efficiency of the eccentricity-driven excavation process on Phobos.} 
\end{figure}

\section{Perspectives}
\label{S:4}
We have shown that Phobos' small eccentricity can drive changes in the surface slope that mobilize grains on the top-layers of high-slope regions. While the rate of mass flow is slow, the accumulated effect over time is enough to expose fresh subsurface material, giving a plausible explanation for the dichotomy between red and blue geologic units on the surface. The analysis described here relies on a simplistic model of space weathering based on other airless bodies, and does not take into account Phobos' unique environment, which is not well understood. However, since the predicted surface refreshing is tied to Phobos' orbital motion, we can place new constraints on the space weathering timescale for Phobos itself, which may then allow us to better understand this process for other airless bodies in our Solar System. With a better understanding of the surface and subsurface conditions on the Martian moon, the Phobos surface can become a ``Rosetta-stone'' for understanding space weathering elsewhere in the solar system. MMX\cite{Kawakatsu2017} (led by JAXA, with contributing instruments from CNES and NASA) will explore Phobos and obtain a core sample from its surface at a depth of up to 10 cm. The combination of a returned sample, new remote sensing observations of the Phobos surface, and a better understanding of surface-refreshing process on Phobos will revolutionize our understanding of the interaction between the surface of an airless body and the space environment. 

Finally, one of MMX's main goals is to provide an answer to the question of Phobos' enigmatic origin. It is unclear whether Phobos is a captured Solar System body \cite{Burns1978,Hansen2018}, or if it formed in an accretion disk following a giant impact with Mars \cite{Craddock1994,Hyodo2017}. Current theories of a giant impact or in-situ formation origin predict that Phobos should be made up of a relatively large proportion of Martian material and imply the presence of mafic minerals such as olivine and  pyroxene. The observed lack of diagnostic absorption features in spectral data suggests that either: a) all of Phobos' surface is heavily space weathered, removing spectral features, b) the surface grains are extremely fine, or c) the regolith is substantially mixed with opaque materials such as carbon. Our analysis suggests that the blue units on Phobos represent pristine endogenic material. Therefore, the lack of mafic absorption features in the spectral data of blue units \cite{Fraeman2012}, which are relatively featureless, would suggest that the first scenario (a) is unlikely, and that it is possible that Phobos has a sparse abundance of Martian material. Therefore, we find that this cold-flow mechanism which can refresh the Phobos surface indicates that Phobos is not spectrally like Mars; or, the observed lack of mafic absorptions must be explained in some other scenario, such as those listed above.

\input{methods}

\input{references}

\input{acknowledgments_contributions}

\end{document}

%% file: abstract.tex
\section*{Abstract}
\textbf{The surface of the Martian moon Phobos exhibits two distinct geologic units, known as the red and blue units. The provenance of these regions is uncertain  yet crucial to understanding the origin of the Martian moon and its interaction with the space environment. Here we show that Phobos' orbital eccentricity can cause sufficient grain motion to refresh its surface, suggesting that space weathering is the likely driver of the dichotomy on the moon's surface. In particular, we predict that blue regions are made up of pristine endogenic material that can be uncovered in steep terrain subject to large variations in the tidal forcing from Mars. The predictions of our model are consistent with current spacecraft observations which show that blue units are found near these regions.}


%% file: methods.tex
\section*{Methods}
\subsection*{The dynamical environment on Phobos' surface}
A point mass in the vicinity of Phobos is subject to the gravitational attraction of both the planetary satellite itself and Mars. Accordingly, the acceleration felt by particle grains on the surface of the Martian moon can be approximated by the equations of the ERTBP \cite{Broucke1969}. Since the distance between the grain and the barycenter of Phobos is much smaller than the distance between the grain and the barycenter of Mars, the equations of motion can be further simplified by expanding the gravitational term of Mars in a Taylor series up to the first order. The resulting system of equations of motion is known as the Hill approximation of the ERTBP and is better understood in a pulsating rotating reference frame centered on Phobos and such that $\versor{x}$ is constantly align with an imaginary line connecting the two primaries, $\versor{z}$ is parallel to the orbital angular momentum of the moon, and $\versor{y} = \versor{z} \times \versor{z}$. Following the length and time scaling described in \cite{Scheeres2002}, the equations read as
\begin{equation}
\left\{
\begin{array}{ccl}
x'' & = & \dfrac{1}{1+e\,\cos{\nu}}\,(g_x + 3\,x) + 2\,y',\\
y'' & = & \dfrac{1}{1+e\,\cos{\nu}}\,g_y - 2\,x',\\
z'' & = & \dfrac{1}{1+e\,\cos{\nu}}\,g_z - z,\end{array}
\right.
\end{equation}
where $(\cdot)'$ denotes differentiation with respect to $\nu$, i.e., the true anomaly of Phobos, $X = \mat x, y, z, x', y', z'\matf^T$ represents the state of the mass particle, $e = 0.0151$ is the eccentricity of Phobos, and $\vect{g} = \mat g_x, g_y, g_z\matf^T$ is the gravitational attraction of the Martian moon. The acceleration is obtained in closed form via a constant density polyhedron model consisting of $40962$ vertices, $81920$ facets, and density equal to $1860$ km/m$^3$\cite{Werner1996, Willner2014, Murchie2015}.

In this paper, we are interested in monitoring the variation of the surface slopes over one orbital period of Phobos around Mars. Accordingly, let $\versor{n}_i$ and $\vect{r}_i$ denote the normal vector and barycenter position vector of the i-th facet, respectively. The surface slope is obtained as the supplement of the angle between $\versor{n}_i$ and the acceleration vector $\vect{a}_i = \mat x’’, y’’ , z’’\matf^T$ evaluated in $\vect{X}_i = \mat x_i, y_i, z_i, 0, 0, 0\matf^T$ at $\nu = \nu^*$, i.e.,
\begin{equation}
\theta(\nu^*) = \pi - \arccos{[(\versor{n}_i \cdot \vect{a}_i) / \| \vect{a}_i \| ]}.
\end{equation}

While Phobos rotates at a constant rate of $\omega \simeq 3.13$ rev/day, the pulsating reference frame rotates about the $\versor{z}$-axis with a time-changing angular velocity of $\dot{\nu} = \omega\,\dfrac{(1 + e\,\cos{\nu})^2}{(1-e^2)^{3/2}}$. Consequently, as seen from the rotating reference frame, the polyhedral shape of Phobos describes librations of approximately $\pm 1$ deg about the $\versor{x}$ axis. These librations cause periodic changes in the tidal acceleration due to Mars that may be resposible for triggering the cold flow mechanism in high-surface slopes regions of Phobos. 

\subsection*{Granular Dynamics Simulation}
For the local simulations of granular flow, we used the \textit{N}-body code \texttt{PKDGRAV} \cite{Richardson2011} which is capable of accurately simulating the complexity of grain-grain and grain-boundary interactions through a soft-sphere discrete element method \cite{Schwartz2012} (SSDEM). In SSDEM, collisions of spherical grains are resolved by allowing them to slightly overlap (typically $< 1$\% of their radii) and then applying multi-contact and multi-frictional forces, including static, rolling and twisting friction. Modeling grain friction accurately is a critical component for high-fidelity granular physics simulations. For \texttt{PKDGRAV}, a new rolling friction model has recently been implemented \cite{Zhang2017} that allows for a more accurate modeling of grain shape and angularity, through a shape parameter $\beta$. 

The mechanical properties of Phobos' surface are poorly known; however, the range of plausible material types are limited, and some remote sensing observations can help constrain some important parameters that are required for accurate simulations. Radar albedo measurements \cite{Busch2007} provide an estimate of the near surface porosity ($40 \pm 10$\%). Local slopes on the surface of Phobos can be as high as 45$^{\circ}$; however, these are likely regions that are in a super-critical state and may not accurately reflect the material's true angle of repose. We chose a set of friction parameters for the grains such that they emulate the static behavior of sand of medium hardness as determined by laboratory experiments\cite{Jiang2015}, which have a corresponding angle of friction of 34$^{\circ}$. The grain size of Phobos regolith is highly uncertain. Early measurements from the Viking spacecraft suggested that the typical grain size of Phobos regolith is 50-100 $\mu$m\cite{Sasaki1995}. More recent work that used thermal inertia measurements combined with a thermal conductivity model for regolith suggests a larger average grain size of 1.1 mm\cite{GundlachBlum2013} ; however, there are large uncertainties ($+0.9/-0.7$ mm). We performed a few high-resolution simulations with grains that have radii of $R_{\mathrm{grain}}$ = 0.9-1.1 mm; however, the combination of particle number (50,000 particles), simulation time (7 h 39.2 m), and small timestep (0.05 s) make the computation very expensive. Therefore, to speed up the calculation, we used a few high-resolution simulations as a baseline, and the majority of our simulations were lower-resolution cases ($R_{\mathrm{grain}}$ = 1.4-1.6 cm, N=7,000). This was necessary to sweep a larger parameter space in starting slope, $\theta$, and slope change amplitude, $\delta \theta$. Table 1 summarizes the properties of grains and regolith bed for our simulations.

\begin{table}[h]
\centering
\begin{tabular}{l l l}
\hline
\textbf{Property} & \textbf{Low Resolution} & \textbf{High Resolution}\\
\hline
$R_{\mathrm{grain}}$ & 1.4-1.6 cm & 0.09-0.11 cm \\
$\phi$ & 34$^{\circ}$ & 34$^{\circ}$ \\
$\rho_{\mathrm{grain}}$ & 3.1 g cm$^{-3}$ & 3.1 g cm$^{-3}$ \\
$\rho_{\mathrm{bulk}}$ & 1.7 g cm$^{-3}$ & 1.7 g cm$^{-3}$ \\
$N$ & 7,000 & 50,000 \\
Sim. Area & 1.5 $\times$ 1.2 m & 0.60 $\times$ 0.4 m \\
\hline
\end{tabular}
\caption{Properties of simulated regolith for granular dynamics simulations. $R_{\mathrm{grain}}$: grain radius, $\phi$: critical angle, $\rho_{\mathrm{grain}}$: grain density, $\rho_{\mathrm{bulk}}$: grain bed bulk density, $N$: number of particles, Sim. Area: surface area of grain bed.  }
\end{table}

Since we are interested in measuring the possibility of surface grain motion due to these small variations in the surface slope, and not complete hillslope-failure, we restrict the vertical scale of our simulations to 5 grain diameters. We set up simulations with initial slopes ranging from 5$^{\circ}$ to 32$^{\circ}$ and amplitudes of variation between 0.5$^{\circ}$ and 1.5$^{\circ}$. This is done by performing an initial setup simulation where the direction of the gravity vector is changed till the desired gravity slope is achieved. The change in the gravity vector is done gradually (over 1 hr in simulated time) as a precaution so as to prevent granular flow from triggering before the time-varying slope change is simulated. The grains are then allowed to settle till all small motions are dissipated. Only after this setup is complete do we perform a full simulation, varying the gravity slope of the inclined granular bed over 1 Phobos orbit (7hr 39 minutes). 

For the limited number of high-resolution cases, we found that, for the same $\theta$ and $\delta \theta$, the particle flux (particle/m$^2$/orbit) was similar to low resolution cases. Therefore, we expect the volume flux (m$^3$/m$^2$/orbit) to depend on the actual grain size on the surface of Phobos. For the same period of time, a surface made up of small particles will excavate material at a given depth at a slower rate than a surface with larger particles. Future work will focus on testing this finding in order to develop scaling laws for the volume flux as a function of grain size. 

While the grains we simulate here are cohesion-less, the actual grains on Phobos are likely to have some non-zero cohesion that will influence the grain dynamics \cite{Scheeres2010}. Cohesive bonding between grains may act as an inhibitor to particle motion, increasing an individual particle's shear strength; however, it is not obvious how grains may actually bond in the Phobos environment. In particular, a size distribution of grains may lead to small grains being bonded to larger ones, and being carried away by the eccentricity-driven mass-wasting process, enhancing the expected excavation rates. Furthermore, as the sizes of the grains become smaller, the relative importance of cohesion is magnified, and processes such as grain lofting\cite{HartzellScheeres2011}, driven by charging from the solar wind or from passing coronal mass ejections\cite{Farrel2017}, may be present on Phobos' surface. Therefore, the contribution of cohesion to this process may be non-trivial, and we will investigate its influence in the future with a recently implemented cohesion-model in \texttt{PKDGRAV}\cite{Zhang2018}.

\subsection*{Comparing excavation rates to space weathering rates}
We calculate the necessary volume flux required to uncover fresh material (Fig.\ 4) by equating an excavation rate with a space weathering rate to a certain depth. We define the excavation rate, $EXR$, to be the volume, $V$, out of an area, $A$, for a given time period, $\delta t$,
\begin{equation}\label{Eq:EXR}
EXR = \frac{V}{A \delta t},
\end{equation}
and we define the space weathering rate, $SWR$, as the timescale,
$t_{\mathrm{SW}}$ necessary to mature regolith up to a depth of $d_{\mathrm{SW}}$.
\begin{equation}\label{Eq:SWR}
SWR = d_{\mathrm{SW}}/t_{\mathrm{SW}}
\end{equation}
We calculate the equilibrium values for excavation and weathering to different depths by equating equations (\ref{Eq:EXR}) and (\ref{Eq:SWR}), setting the area to units of 1m$^2$, $\delta t$ to one Phobos orbit, and the grain radius, $R_{\mathrm{grain}}$ to the volume of a mm-size particle, such that the number
of particles that must flow out of a 1m$^2$ region in 1 Phobos orbital period, $T_{\mathrm{Phobos}}$, to excavate that region to a depth of $d_{\mathrm{SW}}$ in $t_{\mathrm{SW}}$ years is given by:
\begin{equation}\label{Eq:QRAte}
N_{mm} = \frac{d_{\mathrm{SW}}}{t_{\mathrm{SW}}} \frac{1\mathrm{m}^2}{A} \frac{\delta t}{T_{\mathrm{Phobos}}} \left(\frac{10^{-3}\mathrm{m}}{R_{\mathrm{grain}}}\right)^3
\end{equation}

\subsection*{Space weathering timescales for Phobos from spectral slopes}

We derive spectral slopes from CRISM spectra of representative blue and red regions on Phobos corrected to I/F with incidence and phase angles equal to 30 degrees and emission angle of 0 degrees \cite{Fraeman2012}. We measure spectral slopes following \cite{Kaluna2016} using the normalized reflectivity gradient S' defined by \cite{JewittMeech1986}:
\begin{equation}
S'(\lambda_1,\lambda_2) = \dfrac{dS/d\lambda}{S_{0.55}}\,
\end{equation}
where $dS/d\lambda$ is the slope of the reflectance between wavelengths $\lambda_1 = 0.49 \mu$m and $\lambda_2 = 0.91 \mu$m and slopes are normalized to reflectance at $0.55 \mu$m. We find $S'_{red}=0.69\%\mu$m$^{-1}$ and $S'_{blue}=0.29\%\mu$m$^{-1}$, for an overall slope change from blue to red of $0.4\%\mu$m$^{-1}$. Slope change rates due to space weathering have been estimated for several asteroid families of different compositions: \cite{Kaluna2016} observed a slope change of $0.08\%\mu$m$^{-1}$ over 2.3 Gyr for C-complex asteroids, while \cite{Lazzarin2006} determined a rate of $8.8\times 10^{-5}$AU$^{2}\%\mu$m$^{-1}$Myr$^{-1}$ for C-complex and $24.9\times 10^{-5}$AU$^{2}\%\mu$m$^{-1}$Myr$^{-1}$ for S-complex asteroids. \cite{Vernazza2009} determined that asteroids may weather relatively quickly, with a slope change of $0.4\%\mu$m$^{-1}$ on the order of a million years. Although the space weathering environment may be very different at Phobos than for asteroids, and Phobos’ composition is not well constrained, these space weathering rates can be used to obtain an order of magnitude of the effect expected for Phobos. Using these slope change rates due to space weathering of asteroid families and the slope difference of $0.4\%\mu$m$^{-1}$ between blue and red regions, we estimate a space weathering age for Phobos’ red regions on the order of $10^{6}-10^{10}$ yr. While the upper time range based on slow C-type asteroid reddening is physically infeasible, it is the lower time limit that is relevant to our study, in order to determine whether the cold flow mechanism is efficient enough to compete with rapid space weathering.

%% file: references.tex

%% file: acknowledgments_contributions.tex
\section*{Acknowledgements}
R.L.B acknowledges support from JAXA's Aerospace Project Research Associate Program. N.B. conducted this work as a JSPS International Research Fellow. S.T.C. was supported by the JAXA International Top Young Fellowship Program. The authors also thank Patrick Michel for constructive feedback on the results and implications of this work. Grain dynamics simulations were calculated on YORP cluster run by the Center for Theory and Computation at the Department of Astronomy at the University of Maryland. For data visualization, the authors made use of the freeware, multi-platform, ray-tracing package, Persistence of Vision Raytracer.

\section*{Author Contributions}
R.L.B. conceptualized the study, designed and performed the local simulations of granular dynamics, and led the research. N.B. initiated the project through a study of the three body elliptical problem on Phobos, performed the gravitational dynamics calculations, and contributed to the analyses. S.T.C. provided geophysical and geomorphological expertise, remote sensing analysis, and constructed the model for regolith development on Phobos. Y.K. provided guidance and advice on the formulation and scope of the research. M.F. provided guidance and discourse on the implications of the results. Y.K. and M.F. provided expertise in small body exploration and contextualized the research in the frame of the MMX mission. All authors contributed to the interpretation of the results and preparation of the manuscript.

\section*{Competing interests}
The authors declare no competing interests.

\section*{Data Availability}
The datasets generated and analyzed during the current study are available from the corresponding author on reasonable request.

\section*{Code Availability}
The code used to generate the datasets are available from the corresponding author on reasonable request.